\documentclass[runningheads]{svmult}

\usepackage{makeidx}   
\usepackage{graphicx}  
\usepackage{subeqnar}  
\usepackage{multicol}  
\usepackage{physprbb}  
\makeindex             

\begin{document}
\title*{Evidence for Core Collapse in the Type Ib/c SN~1999ex}
\toctitle{Evidence for Core Collapse in the Type Ib/c SN~1999ex}
%
%
\titlerunning{The Type Ib/c SN~1999ex}
%
\author{Mario Hamuy\inst{1}
\and Maximilian Stritzinger\inst{2}
\and M. M. Phillips\inst{1}
\and Nicholas B. Suntzeff\inst{3}
\and Jos\'{e} Maza\inst{4}
\and Philip A. Pinto\inst{5} }
\authorrunning{Mario Hamuy et al.}
%
%
\institute{Carnegie Observatories, 813 Santa Barbara Street, Pasadena, CA 91101, USA
\and Department of Physics, The University of Arizona, Tucson, AZ 85721, USA
\and Cerro Tololo Inter-American Observatory, Casilla 603, La Serena, Chile
\and Departamento de Astronom\'\i a, Universidad de Chile, Casilla 36-D, Santiago, Chile
\and Steward Observatory, The University of Arizona, Tucson, AZ 85721, USA}

\maketitle              

\begin{abstract}
We present optical and infrared spectra of SN~1999ex, which are
characterized by the lack of strong hydrogen lines, weak optical He I lines,
and strong He I $\lambda$10830,20581. SN~1999ex provides a clear example of
an intermediate case between pure Ib and Ic supernovae, which suggests a
continuous spectroscopic sequence between SNe~Ic to SNe~Ib. Our $UBVRIz$
photometric observations of SN~1999ex started only one day after explosion,
which permitted us to witness an elusive transient cooling phase that lasted 4 days.
The initial cooling and subsequent heating due to $^{56}$Ni$\rightarrow$$^{56}$Co$\rightarrow$$^{56}$Fe
produced a dip in the lightcurve which is consistent with explosion models involving core collapse of evolved
massive helium stars, and not consistent with lightcurves resulting from the
thermonuclear runaway of compact white dwarfs.
\end{abstract}

\section{Introduction}
In a rare occurrence the spiral galaxy IC~5179 ($cz$=3,498 $km$ $s^{-1}$) produced two
supernovae (SNe) in an interval of only three weeks. The first object (SN~1999ee) was a Type Ia event
discovered by us 10 days before maximum \cite{maza99}. The early discovery motivated us to use the YALO
and 0.9-m telescopes at the Cerro Tololo Inter-American Observatory in order to secure nightly
$UBVRIz$ photometric observations of this event, and the YALO and Las Campanas 1-m and 2.5-m telescopes
to obtain $JHK$ photometry. Although the second object (SN~1999ex) exploded three
weeks later and was promptly present in our CCD images we did not notice its presence.
Its discovery had to await independent observations obtained at Perth Observatory \cite{martin99}.
Once SN~1999ex was reported to the IAU Circulars we initiated an optical and
infrared (IR) spectroscopic followup using the European Southern Observatory  NTT
and Danish 1.5-m telescope at La Silla, and the VLT at Cerro Paranal. Our spectroscopic
and photometric observations of SN~1999ex constitute an unprecedented dataset which
provides support to our understanding of the nature of core collapse SNe.
In this paper we show some of our observations and their interpretation.
For a detailed report of our observations the reader is referred to \cite{hamuy02},
\cite{stritzinger02}, and \cite{krisciunas02}.

\section{Spectroscopic Observations}

\begin{figure}[b]
\begin{center}
\includegraphics[width=.6\textwidth]{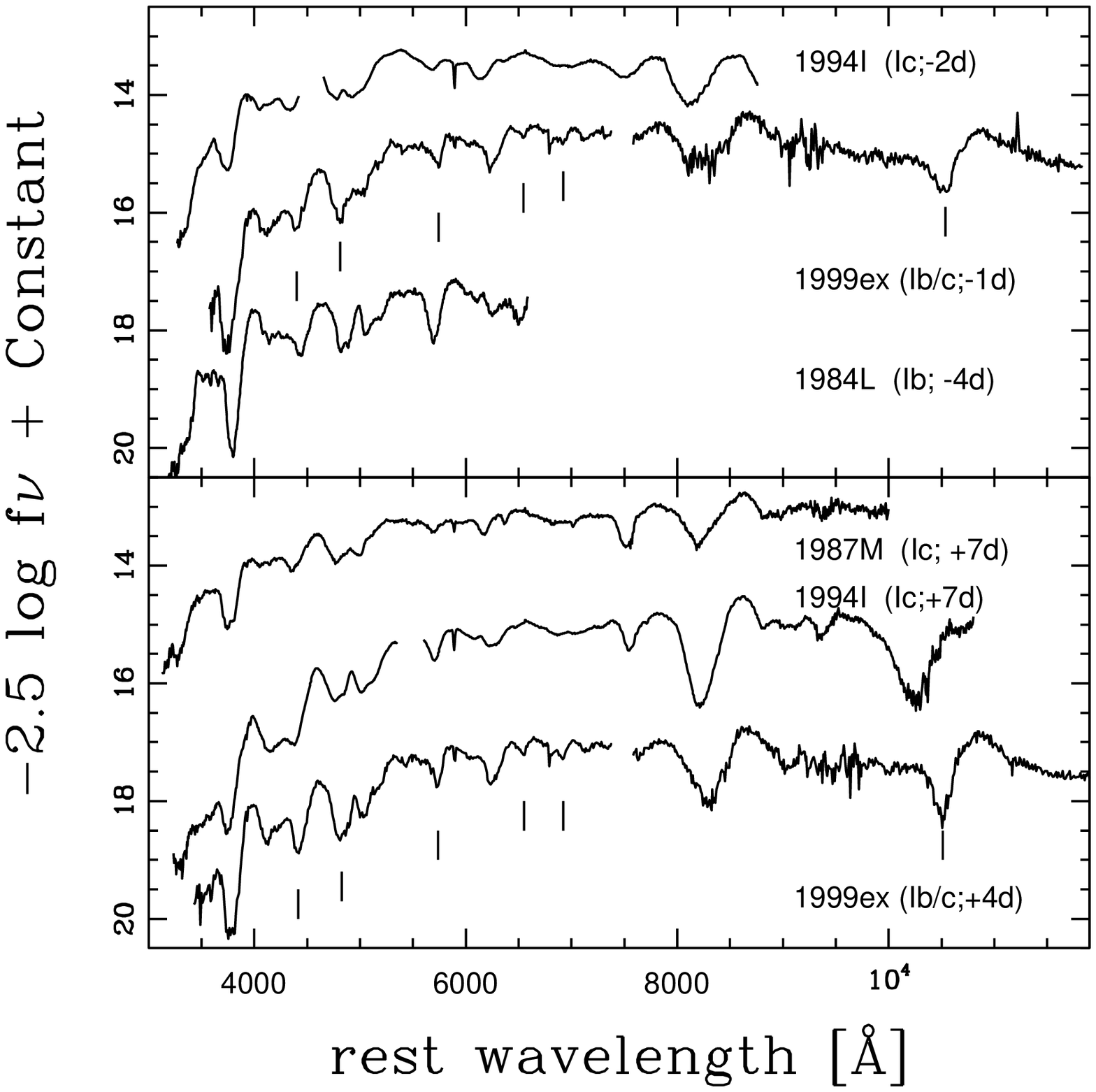}
\end{center}
\caption[]{Comparison of spectra of the Type Ib/c SN~1999ex with the prototype
of the Ib class SN~1984L \cite{harkness87}, and the Type Ic SNe~1994I \cite{filippenko95} and 1987M \cite{filippenko90}.
{\it Tick marks} indicate the He I lines in the SN~1999ex spectra.
The strengths of the helium lines gradually increase from the
Type Ic to the Ib SN, and SN~1999ex provides a link 
between these two subclasses. Time (in days) since $B$ maximum is indicated for each spectrum}
\label{hamuyF1.fig}
\end{figure}

Fig.~\ref{hamuyF1.fig} compares the near-maximum optical spectra of SN~1999ex to those of
the prototype of the Ib class SN~1984L \cite{harkness87}, and the Type Ic SNe~1994I
\cite{filippenko95} and 1987M \cite{filippenko90}. The first spectrum of SN~1999ex
was characterized by a reddish continuum and several broad absorption/emission
features due to Ca II H\&K $\lambda\lambda$3934,3968, Na I D $\lambda\lambda$5890,5896,
and the Ca triplet with a clear P-Cygni profile. This spectrum bore quite resemblance
to that of the Type Ic SN~1994I \cite{filippenko95}. However, SN~1999ex showed
evidence for He I absorptions (shown with tick marks) of moderate strength in the optical region,
thus suggesting the existence of an intermediate Ib/c case. Our observations
of SN~1999ex provide a clear link between the Ib and Ic classes and suggest
that there is a continuous sequence of SNe Ib and Ic objects.

\begin{figure}[b]
\begin{center}
\includegraphics[width=.6\textwidth]{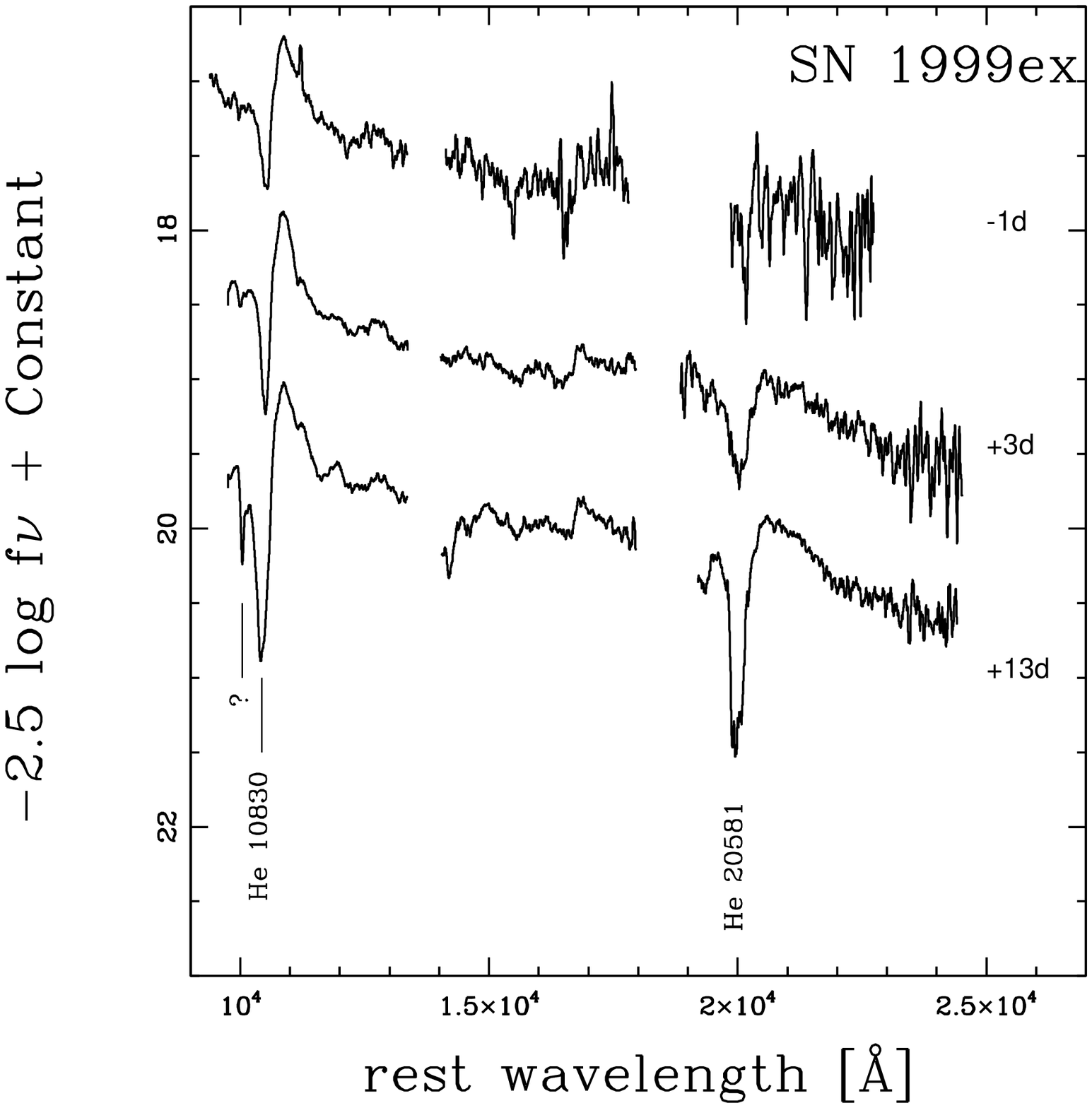}
\end{center}
\caption[]{IR spectroscopic evolution of SN~1999ex.
The two most prominent features are attributed to He I.
Time (in days) since $B$ maximum is
indicated for each spectrum}
\label{hamuyF2.fig}
\end{figure}

The presence of helium in the atmosphere of SN~1999ex can be further examined in
our IR observations shown in Fig. \ref{hamuyF2.fig}. The strong feature
near 1.05 $\mu$m probably had a significant contribution from He I $\lambda$10830,
although it could be blended with lines of C~I and Si~I \cite{millard99,baron99}.
If this feature was He I $\lambda$10830 it would imply an expansion velocity of
6,000-8,000 km~s$^{-1}$, which matches very well the velocities derived from the Fe and Na lines. 
This identification is supported by the presence of He I $\lambda$20581 
with the same expansion velocity deduced from He I $\lambda$10830.
Our IR spectra of SN~1999ex provide unambiguous proof that helium was present
in the atmosphere of this intermediate Ib/c object. A detailed atmosphere
model could be very useful at placing specific limits on the helium mass in the
ejecta of SN~1999ex and constraining the nature of its progenitor.
Evidently, $K$-band spectroscopy is probably the best tool to
explore the presence or absence of helium in the atmospheres of SNe~Ib and Ic.

The question of whether or not SNe~Ic have helium is still controversial.
A detailed inspection of the spectra of SN~1994I
led Filippenko et al. \cite{filippenko95} to conclude that weak He I lines were probably present in the
optical region and that He I $\lambda$10830 was very prominent, although its Doppler
shift implied an unusually high expansion velocity $\sim$16,600 km~s$^{-1}$ as can be
seen in the bottom panel of Fig. \ref{hamuyF1.fig}.
Clocchiatti et al. \cite{clocchiatti96} confirmed these observations and found evidence that He I $\lambda$5876
was also present in SN~1994I with a blueshift of $\sim$17,800 km~s$^{-1}$. They also reported
high velocity He I $\lambda$5876 in the spectra of the three best-observed Type Ic SNe (1983V, 1987M, and 1988L),
which led them to conclude that most, and probably all, SNe~Ic have optical He I lines.
This conclusion, however, was recently questioned by Millard et al. \cite{millard99} and Baron et al. \cite{baron99} by means of
spectral synthesis models which showed that the 1.05 $\mu$m feature could
be accounted with lines of C I and Si I. Moreover, Baron et al. \cite{baron99} argued that
the feature attributed to He I $\lambda$5876 in the spectrum of SN~1994I could
be a blend of other species. Recently, Matheson et al. \cite{matheson01} compiled and analyzed a large
collection of spectra of SNe~Ib and Ic and did not find compelling evidence for helium
in the spectra of SNe~Ic. Evidently, there is no
consensus yet about this issue.
The existence of an intermediate Ib/c object such as SN~1999ex suggests a
continous transition between SNe Ib and SNe~Ic so it is likely that some
SNe~Ic have some helium in their atmospheres.

\section{Photometric Observations}

\begin{figure}[b]
\begin{center}
\includegraphics[width=.6\textwidth]{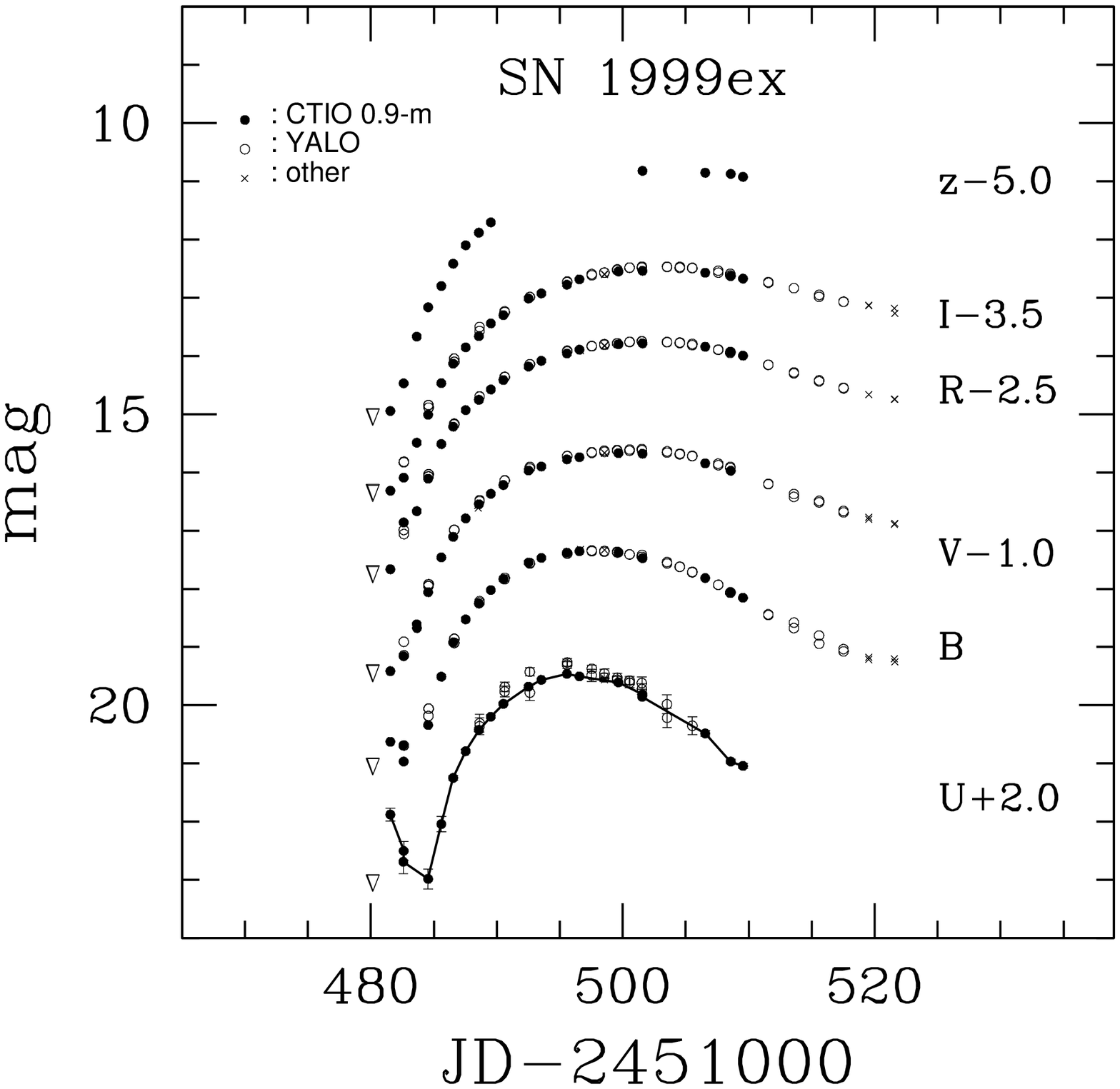}
\end{center}
\caption[]{$UBVRIz$ lightcurves of SN~1999ex measured with the YALO ({\it open points})
and CTIO 0.91-m telescopes ({\it closed points}). Upper limits derived from images taken
on JD 2,451,480.5 are also included ({\it open triangles}). The {\it solid line} through
the $U$ data is drawn to help the eye to see the initial upturn}
\label{hamuyF3.fig}
\end{figure}

Fig.~\ref{hamuyF3.fig} shows the $UBVRIz$ lightcurves of SN~1999ex. Clearly the observations
began well before maximum light thanks to our continuous followup of IC~5179 owing to
the prior discovery of SN~1999ee. The first detection occurred on JD 2,451,481.6 in all filters.
Excellent seeing images obtained on the previous night allowed us to place reliable upper limits to the SN
brightness, which permitted us to conclude that the explosion took place on JD 2,451,480.5 ($\pm$0.5).
Along with the Type Ic hypernova 1998bw \cite{galama98}, these are the earliest observations of a 
SN~Ib/c. The most remarkable feature in this figure is the early dip in the $U$ and $B$
lightcurves -- covering the first 4 days of evolution -- after which the SN steadily rose
to maximum light. Similar initial upturns have been observed before in SN~Ic~1998bw \cite{galama98},
SN~II~1987A \cite{hamuy88}, and SN~IIb~1993J \cite{schmidt93,richmond94}.
For these SNe~II it is thought that the initial dip corresponded to a phase of adiabatic cooling
that ensued the initial UV flash caused by shock emergence which super-heated and accelerated the photosphere.
The following brightening is attributed to the energy deposited behind the photosphere by the radioactive
decay of $^{56}$Ni$\rightarrow$$^{56}$Co$\rightarrow$$^{56}$Fe. This similarity suggests that
the progenitor of SN~1999ex was a massive progenitor that underwent core collapse.
Woosley et al. \cite{woosley87} computed Type Ib SN models consisting of the explosion
of an evolved 6.2 M$_\odot$ helium star. Their Figure 7 shows the bolometric
luminosities of three models with different explosion energies and
$^{56}$Ni nucleosynthesis, all of which show an initial peak followed by
a dip a few days later and the subsequent brightening caused by
$^{56}$Ni $\rightarrow$ $^{56}$Co $\rightarrow$ $^{56}$Fe, making them
good models for SN~1999ex.

\begin{figure}[b]
\begin{center}
\includegraphics[width=.6\textwidth]{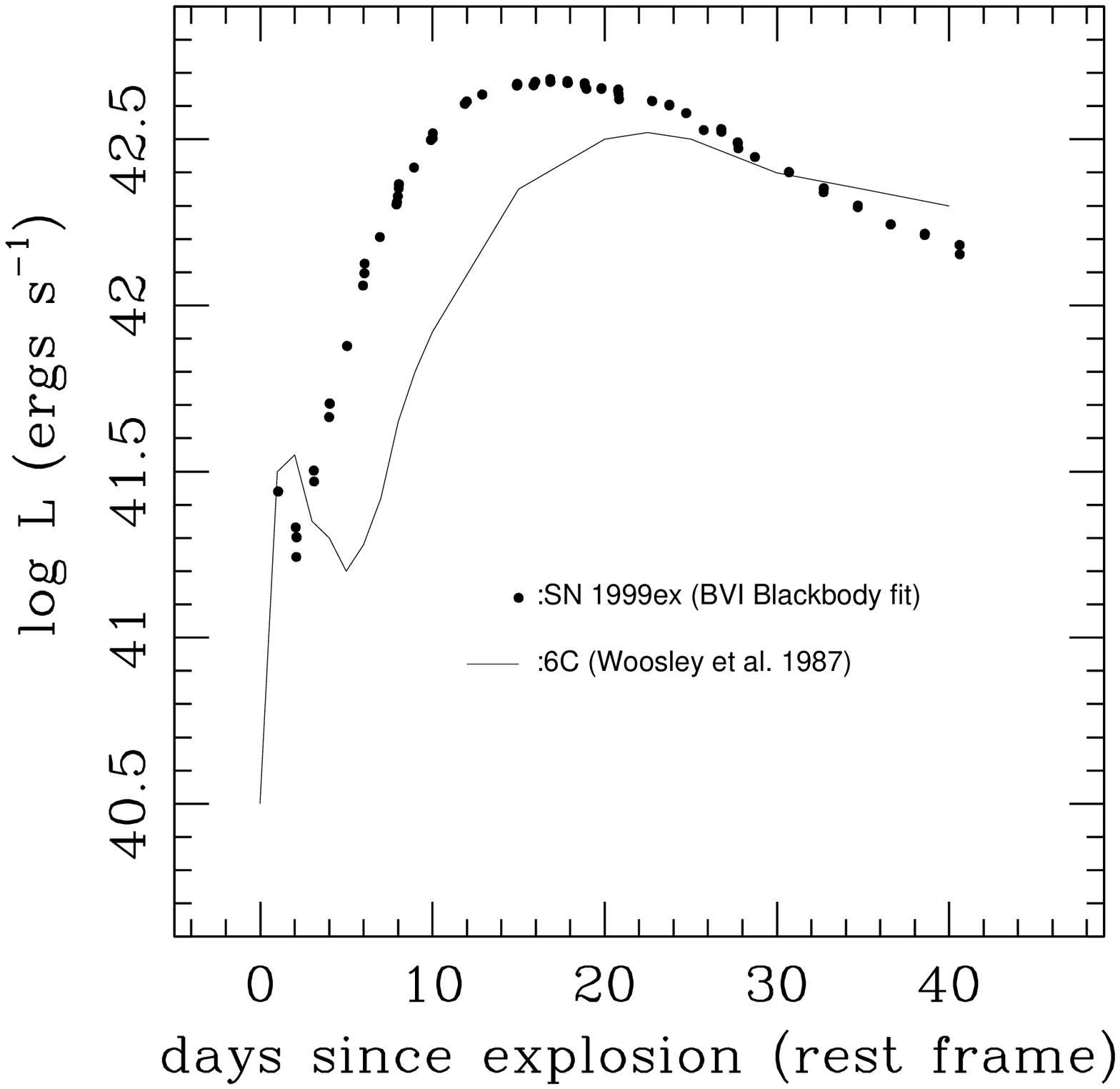}
\end{center}
\caption[]{Bolometric lightcurve of SN~1999ex derived from blackbody fits to the $BVI$
magnitudes, assuming $E(B-V)$$_{host}$=0.28 and a distance of 51.2 Mpc ({\it closed circles}).
The {\it solid line} is the 6C hydrogenless core bounce SN model of Woosley et al. \cite{woosley87}}
\label{hamuyF4.fig}
\end{figure}

In order to compare the observations with the models we computed a bolometric lightcurve
for SN~1999ex by performing blackbody (BB) fits to our $BVI$ photometry (Fig.~\ref{hamuyF4.fig})
Among the three SN~Ib models of Woosley et al., the one 
with kinetic energy of 2.7$\times$10$^{51}$ ergs and 0.16 M$_\odot$ of $^{56}$Ni
provides the best match to SN~1999ex. The agreement is remarkable
considering that we are not attempting to adjust the parameters.
The initial peak and subsequent dip have approximately the right luminosities
although the evolution of SN~1999ex was somewhat faster. The following
evolution is well described by the model.

The observation of the tail of the shock wave breakout in SN~1999ex and the initial dip in
the lightcurve provides us with an insight on the
type of progenitor system for SNe~Ib/c. Several different models have been
proposed as progenitors for this type of SNe. One possibility is
an accreting white dwarf which may explode via thermal detonation
upon reaching the Chandrasekhar mass \cite{sramek84,branch86}.
These models are expected to produce lightcurves with an initial peak that corresponds to the
emergence of the burning front, a fast luminosity drop due to adiabatic expansion,
and a subsequent rise caused by radioactive heating. Given the compact nature of
the progenitor ($\sim$1,800 km) the cooling time scale by adiabatic expansion
is only a few minutes \cite{hoflich02} and the lightcurve
is entirely governed by radioactive heating \cite{hoflich96}. Hence,
these models are not expected to show an early dip at a few days past
explosion as is observed in SN~1999ex.
The second and more favored model for SNe~Ib/c consists of core collapse of
massive stars ($M_{ZAMS}$$>$8 M$_{\odot}$) which lose their outer H envelope before
explosion. Within the core collapse models, there are two basic types
of progenitor systems: 1) a massive ($M_{ZAMS}$$>$35 M$_{\odot}$) star which undergoes strong
stellar winds and becomes a Wolf-Rayet star at the time of explosion \cite{woosley93},
and 2) an exchanging binary system \cite{shigeyama90,nomoto94,iwamoto94} for less massive stars.
The resulting SNe have lightcurves containing an initial spike followed by a dip.
Since the initial radii of these progenitors are $\sim$100 times greater than that of white
dwarfs, the dip occurs several days after explosion \cite{woosley87,shigeyama90,woosley93}, very much like SN~1999ex.

Although a detailed modeling of SN~1999ex is beyond the scope of this paper,
our fortuitous observations lend support to the idea that SNe~Ib/c are due to core collapse of
massive progenitors rather than thermonuclear disruption of white dwarfs.\\

\noindent M.H. acknowledges support provided by NASA through Hubble Fellowship grant HST-HF-01139.01-A
awarded by the Space Telescope Science Institute.

%

\end{document}